\documentclass[twocolumn, prb]{revtex4}
\usepackage{graphicx}
\usepackage{amsmath}
\usepackage{color}
\usepackage{array}
\DeclareGraphicsExtensions{.pdf, .jpg, .png, .eps}
\newcommand {\NF}{N_{\rm F}}
\newcommand {\EF}{\epsilon_F}

\newcommand {\bbk}{\mathbf k}
\newcommand {\bbq}{\mathbf q}
\newcommand {\bbkk}{\mathbf {k'}}

\newcommand {\ek}{\epsilon_{\mathbf k}}
\newcommand {\ekk}{\epsilon_{\mathbf {k'}}}
\newcommand {\oql}{\omega_{\mathbf q \nu}}

\newcommand {\on}{\omega_n}
\newcommand {\onp}{\omega_{n'}}

\newcommand {\afkko}{\alpha^{2} F(\mathbf k,\mathbf {k'},\omega)}

\newcommand {\afo}{\alpha^{2} F(\omega)}

\newcommand {\gkkl}{g_{\mathbf k \mathbf {k'} \nu}}

\begin{document}
          
\title{Anisotropic Migdal-Eliashberg theory using Wannier functions}

\author{E. R. Margine and F. Giustino}

\affiliation{Department of Materials, University of Oxford, Parks Road,
Oxford OX1 3PH, United Kingdom}

\begin{abstract}
We combine the fully anisotropic Migdal-Eliashberg theory with electron-phonon 
interpolation based on maximally-localized Wannier functions, in order to 
perform reliable and highly accurate calculations of the anisotropic  
temperature-dependent superconducting gap and critical temperature of 
conventional superconductors. Compared with the widely used McMillan 
approximation, our methodology yields a more comprehensive and detailed 
description of superconducting properties, and is especially relevant for the 
study of layered or low-dimensional systems as well as systems with complex 
Fermi surfaces. In order to validate our method we perform calculations on two 
prototypical superconductors, Pb and MgB$_2$, and obtain good agreement with 
previous studies.
\end{abstract}

\maketitle

\section{INTRODUCTION}\label{section.introduction}

The prediction of superconducting properties such as the critical temperature 
or the superconducting energy gap remains one of the outstanding challenges 
in modern electronic structure theory.~\cite{carbotte,allen_mitrovic,marsiglio_book,scalapino66,scalapino,
scdft1,scdft2} Owing to the complex nature of the superconducting state, a 
quantitative understanding of the pairing mechanism in conventional 
superconductors requires a very detailed knowledge of the electronic 
structure, the phonon dispersions, and the interaction between electrons 
and phonons. In conventional superconductors below the critical temperature 
electron pairing results from a subtle interplay between the repulsive 
Coulomb interaction and the attractive electron-phonon interaction. Starting 
from the seminal work of Bardeen, Cooper, and Schrieffer (BCS)~\cite{bcs} 
several approaches to the calculation of the superconducting properties have 
been proposed, ranging from semi-empirical methods such as the McMillan 
formula,~\cite{mcmillan} to first-principles Green's function methods such as 
the Migdal-Eliashberg (ME) formalism,~\cite{migdal,eliashberg} and more 
recently also methods based on the density-functional theory concept, such as 
the density functional theory for superconductors (SCDFT).~\cite{scdft1,scdft2}

The vast majority of current investigations rely on the semi-empirical 
McMillan's approach.~\cite{mcmillan} In this approach the entire physics of 
the electron-phonon interaction is condensed into a single parameter
called the electron-phonon coupling strength $\lambda$. The McMillan's method 
works reasonably well for conventional bulk metals and for anisotropic 
superconductors where the Fermi surface anisotropy is smeared out by 
impurities.~\cite{allen_mitrovic,marsiglio_book} However, for layered systems, 
systems of reduced dimensionality, and those with complex multi-sheet Fermi 
surfaces, a careful description of the pairing interactions is crucial and a 
proper treatment of the anisotropic electron-phonon interaction is required. 
The necessity of an anisotropic theory has clearly been demonstrated in two 
important cases such as magnesium diboride, MgB$_2$, and the graphite 
intercalation compound CaC$_6$, using either the ME formalism or the 
SCDFT.~\cite{floris_mgb2,choi_mgb2,sanna07,golubov,mazin,choi_2band,sanna12} 

Unfortunately, the lack of adequate computational tools prevents the research 
community from systematically exploring the importance of anisotropy in 
existing and well characterized superconductors, and for validating 
computational predictions based on the semi-empirical McMillan 
equation.~\cite{giustino_graphane,kolmogorov_lib1,kolmogorov_lib2,kolmogorov_feb} This 
latter aspect is particularly relevant in view of the increasingly important 
role that high-throughput materials design approaches are acquiring in the 
community.~\cite{ceder,curtarolo}

The momentum-resolved superconducting gap and the quasiparticle density of 
states near the Fermi surface can now be measured with unprecedented accuracy 
using high-resolution angle-resolved photoemission spectroscopy~\cite{arpes} 
and scanning tunneling spectroscopy~\cite{sts} experiments. In this context, 
the anisotropic Migdal-Eliashberg formalism promises to be particularly useful
for performing direct comparisons between theory and experiment, and for 
helping establish unambiguously the symmetry of the order parameter.

One critical point which arises when attempting to solve the Eliashberg 
equations of the ME theory or the Bogoliubov-de Gennes equations of the SCDFT 
is that both sets of equations suffer from a strong sensitivity to the 
sampling of the electron-phonon scattering processes in the vicinity of the 
Fermi surface.~\cite{giustino_wannier} The practical consequence of this 
sensitivity is that, in order to achieve numerical convergence, the 
electron-phonon matrix elements must be evaluated for extremely dense electron 
and phonon meshes in the Brillouin zone. This long-standing difficulty was 
overcome in Ref.~\onlinecite{giustino_wannier} by developing an efficient 
first-principles interpolation technique based on maximally-localized Wannier 
functions (MLWF).~\cite{marzari2012} In this method one takes advantage 
of the spatial localization of both electron and phonon Wannier functions in 
order to evaluate only a small number of electron-phonon matrix elements in 
the Wannier representation. These matrix elements are subsequently 
interpolated to arbitrary electron and phonon wavevectors in the Bloch 
representation using a generalized Fourier transform.~\cite{giustino_wannier} 
This method carries general validity and has been demonstrated in several 
other areas, from Fermi surface calculations,~\cite{yates} to 
the anomalous Hall effect,\cite{souza} and more recently for GW 
calculations.~\cite{vanderbilt} A detailed introduction on Wannier-based 
interpolation methods can be found in  Ref.~\onlinecite{marzari2012}.
The scheme of Ref.~\onlinecite{giustino_wannier} is adopted in the present 
work since it provides a robust and efficient framework for developing an 
algorithm to solve the anisotropic Eliashberg equations. Our current 
implementation enables the calculation of the momentum- and band-resolved 
superconducting gap using a very fine Brillouin zone sampling. Without our 
Wannier-based electron-phonon interpolation this operation would not be 
possible owing to the prohibitive computational cost. 

This manuscript is organized as follows. In Sec.~\ref{section.formalism} we 
review the Migdal-Eliashberg theory of superconductivity. A description of the 
computational techniques underpinning the electron-phonon interpolation 
implemented in the {\tt EPW} code~\cite{EPW} is given in 
Sec.~\ref{section.methodology}. In Sec.~\ref{section.applications} we report 
the numerical solutions of the Eliashberg equations for two prototypical 
superconductors, Pb and MgB$_2$. Finally we present our conclusions and outlook
in Sec.~\ref{section.conclusions}. 

\section{MIGDAL-ELIASHBERG FORMALISM}\label{section.formalism}

\subsection{General theory}\label{sec.theory}

A quantitative theory of the superconducting energy gap can conveniently be 
formulated within the framework of the Nambu-Gor'kov 
formalism.~\cite{nambu,gorkov} In this formalism one introduces a 
two-component field operator:
\begin{equation}
\mathbf{\Psi}_{\bbk} =  
   \left( \begin{array}{l}
             c_{\bbk \uparrow} \\ c^\dagger_{-\bbk \downarrow} 
          \end{array} \right).
\label{psi_operators}
\end{equation}
The component $c_{\bbk \uparrow}$ ($c^\dagger_{-\bbk \downarrow}$) of the 
operator destroys (creates) an electron in a Bloch state of combined band and 
momentum index $\bbk$ ($-\bbk$) and spin up (down). A generalized 2$\times$2 
matrix Green's functions $\hat{G}$ is then introduced in order to describe 
electron quasiparticles and Cooper's pairs on an equal 
footing:~\cite{allen_mitrovic,scalapino66}
\begin{equation} \label{eq.G2}
\hat{G}(\bbk,\tau)
= -\langle T_{\tau} \mathbf{\Psi}_{\bbk}(\tau) \mathbf{\Psi}^\dagger_{\bbk}(0) 
\rangle,
\end{equation}
where $T_{\tau}$ is Wick's time-ordering operator for the imaginary time $\tau$
and $\mathbf{\Psi}_{\bbk}(\tau)$ is obtained from Eq.~(\ref{psi_operators}) 
using the Heisenberg picture. The braces indicate a grand-canonical 
thermodynamic average. By replacing Eq.~(\ref{psi_operators}) inside 
Eq.~(\ref{eq.G2}) we find:
\begin{equation}\label{G}
\hat{G}(\bbk,\tau) = -\left[ \begin{array}{cc}
               \langle T_{\tau} c_{\bbk \uparrow}(\tau)
                                c^\dagger_{\bbk \uparrow}(0) \rangle &
               \langle T_{\tau} c_{\bbk \uparrow}(\tau)
                                c_{-\bbk \downarrow}(0) \rangle \\
               \langle T_{\tau} c^\dagger_{-\bbk \downarrow}(\tau)
                                c^\dagger_{\bbk \uparrow}(0) \rangle &
               \langle T_{\tau} c^\dagger_{-\bbk \downarrow}(\tau)
                                c_{-\bbk \downarrow}(0) \rangle
            \end{array} \right].
\end{equation}
Here the diagonal elements correspond to the standard Green's functions 
for electron quasiparticles and describe the dynamics of single-particle 
electronic excitations in the material. On the other hand, the off-diagonal 
elements represent Gor'kov's anomalous Green's functions $F(\bbk,\tau)$ and 
$F^*(\bbk,\tau)$. These functions describe the dynamics of Cooper's pairs and 
are related to the superconducting energy 
gap.~\cite{allen_mitrovic,scalapino66,scalapino} The off-diagonal elements of 
the generalized Green's function in Eq.~(\ref{G}) become nonzero only below 
the critical temperature $T_{\rm c}$, marking the transition to the 
superconducting state.

The generalized Green's function $\hat{G}(\bbk,\tau)$ is periodic in imaginary 
time, therefore it can be expanded using a Fourier series as follows:
\begin{equation}
\hat{G}(\bbk,\tau) =  T \sum_{i\on} e^{-i\on \tau} \hat{G}(\bbk,i\on),
\label{eq.G}
\end{equation}
where $i\on=i(2n+1)\pi T$ ($n$ integer) stands for the fermion Matsubara 
frequencies, and $T$ is the absolute temperature. We use atomic units 
throughout the manuscript, therefore we set $\hbar=k_B=1$. Following 
Eq.~(\ref{eq.G}) the matrix elements of the generalized Green's function read:
\begin{equation} 
\hat{G}(\bbk,i\on) = 
    \left[ \begin{array}{cc}
              G(\bbk,i\on) & F(\bbk,i\on) \\ 
              F^{*}(\bbk,i\on) & -G(-\bbk,-i\on)                        
           \end{array} 
    \right].
\label{G_mat}    
\end{equation}
The study of the superconducting state involves the determination of the 
matrix Green's function in Eq.~(\ref{G_mat}). This can be achieved using 
Dyson's equation:
\begin{equation} 
\hat{G}^{-1}(\bbk,i\on) = 
\hat{G}^{-1}_0(\bbk, i\on) - \hat{\Sigma}(\bbk,i\on),
\label{Dyson_eq}
\end{equation}
where $\hat{G}_0(\bbk, i\on)$ is the electron Green's function for the normal 
state and $\hat{\Sigma}(\bbk,i\on)$ is the self-energy associated with the 
pairing interaction. The normal-state Green's function is calculated by using 
the Kohn-Sham states from density-functional theory to represent 
single-particle excitations. If we denote by $\ek$ the Kohn-Sham eigenvalues 
measured with respect to the chemical potential, and we introduce the Pauli matrices:
\begin{equation}
\begin{array}{ll}
\hat{\tau}_0 = \left( \begin{array}{rr}
                            1 & 0 \\
                            0 & 1                     
                        \end{array} \right), \qquad &
\hat{\tau}_1 = \left( \begin{array}{rr}
                            0 & 1 \\
                            1 & 0                     
                        \end{array} \right), \\
\\                         
\hat{\tau}_2 = \left( \begin{array}{rr}
                            0 & -i \\
                            i & 0                     
                        \end{array} \right), \qquad &
\hat{\tau}_3 = \left( \begin{array}{rr}
                            1 & 0 \\
                            0 & -1                     
                        \end{array} \right),                                   
\end{array}
\label{pauli_mat}
\end{equation}
then the normal-state matrix Green's function acquires the familiar form:
\begin{equation} 
\hat{G}^{-1}_0(\bbk,i\on) = 
 i\on \hat{\tau}_0 - \ek \hat{\tau}_3.
\label{G0}
\end{equation}
Within the Migdal-Eliashberg approximation the electron self-energy leading to 
the superconducting pairing consists of two terms, an  electron-phonon 
contribution $\hat{\Sigma}_{\rm ep}(\bbk,i\on)$ and a Coulomb contribution 
$\hat{\Sigma}_{\rm c}(\bbk,i\on)$:\cite{allen_mitrovic}
\begin{eqnarray} 
\hat{\Sigma}(\bbk,i\on) &=& \hat{\Sigma}_{\rm ep}(\bbk,i\on) + 
\hat{\Sigma}_{\rm c}(\bbk,i\on),
\label{Sigma_1}
\end{eqnarray}
with
\begin{eqnarray} 
\hat{\Sigma}_{\rm ep}(\bbk,i\on) &=& 
   -T \sum_{\bbkk n'} \hat{\tau}_3 \hat{G}(\bbkk,i\onp) \hat{\tau}_3  
   \nonumber \\
   && \hspace{-1.5cm}\times \sum_{\lambda} |\gkkl|^2 
   D_{\nu}(\bbk-\bbkk,i\on-i\onp), 
\label{Sigma_ep}
\end{eqnarray}
and
\begin{equation} 
\hat{\Sigma}_{\rm c}(\bbk,i\on) = 
   -T \sum_{\bbkk n'} \hat{\tau}_3 \hat{G}^{\rm od}(\bbkk,i\onp) \hat{\tau}_3 
   V(\bbk-\bbkk).         
\label{Sigma_c}
\end{equation}
In Eq.~(\ref{Sigma_ep}) 
$D_{\nu}(\bbq,i\on)=2\oql/\left[(i\on)^2-\oql^2\right]$ is the 
dressed propagator for phonons with momentum $\bbq$ and branch index $\nu$, 
and $\gkkl$ is the screened electron-phonon matrix element for the scattering 
between the electronic states $\bbk$ and $\bbkk$ through a phonon with 
wavevector $\bbq\!=\!\bbkk\!-\bbk$, frequency $\oql$ and branch index $\nu$. 
In Eq.~(\ref{Sigma_c}) the $V(\bbk-\bbkk)$'s represent the matrix elements of 
the static screened Coulomb interaction between the electronic states $\bbk$ 
and $\bbkk$.

In writing the electron self-energy of Eqs.~(\ref{Sigma_1})-(\ref{Sigma_c}) 
the following approximations are made: (i) only the first term in the 
Feynman's expansion of the self-energy in terms of electron-phonon diagrams is 
included. This approximation corresponds to Migdal's theorem~\cite{migdal} and 
is based on the observation that the neglected terms are of the order $(m_{\rm 
e}/M)^{1/2}$ ($m_{\rm e}$ being the electron mass and $M$ a characteristic 
nuclear mass). (ii) Only the off-diagonal contributions of the Coulomb 
self-energy are retained. This is done in order to avoid double counting the 
Coulomb effects that are already included in 
$\hat{G}_0(\bbk,i\on)$.~\cite{allen_mitrovic} (iii) The self-energy is 
assumed to be diagonal in the electron band index.~\cite{pickett82} This 
should constitute a reasonable approximation for non-degenerate bands since 
the energy involved in the superconducting pairing is very small, therefore 
band mixing is not expected.

In the literature on the theory of superconductivity it is common practice to 
decompose the matrix self-energy $\hat{\Sigma}(\bbk,i\on)$ in a linear 
combination of Pauli matrices with three scalar functions as coefficients. The 
scalar functions are the mass renormalization function $Z(\bbk,i\on)$, the 
energy shift $\chi(\bbk,i\on)$, and the order parameter $\phi(\bbk,i\on)$, and 
the decomposition reads:
\begin{eqnarray} 
\hat{\Sigma}(\bbk,i\on) &=&
      i\on \left[1-Z(\bbk,i\on) \right] \hat{\tau}_0 
      + \chi(\bbk,i\on) \hat{\tau}_3 \nonumber \\
  &+& \phi(\bbk,i\on) \hat{\tau}_1.
\label{Sigma_2}
\end{eqnarray}
We note that here we have chosen the gauge where the order parameter 
$\overline{\phi}$ is set to zero.~\cite{allen_mitrovic} By replacing 
Eqs.~(\ref{G0}), (\ref{Sigma_2}) inside Eq.~(\ref{Dyson_eq}) 
and solving for the matrix Green's function we obtain:
\begin{eqnarray}
\hat{G}(\bbk,i\on) &=& -   
    \left\{ i\on Z(\bbk,i\on) \hat{\tau}_0
           +\left[ \ek + \chi(\bbk,i\on) \right] \hat{\tau}_3 
    \right. \nonumber \\
&& +\left. \phi(\bbk,i\on) \hat{\tau}_1 \right\} / \Theta(\bbk,i\on),
\label{G_1}
\end{eqnarray}
where the denominator is defined as:
\begin{eqnarray}
\Theta(\bbk,i\on) &=& 
      \left[ \on Z(\bbk,i\on) \right]^2 
     +\left[ \ek + \chi(\bbk,i\on) \right]^2 \nonumber \\ 
&& + \left[ \phi(\bbk,i\on) \right]^2.
\end{eqnarray}
At this point the strategy is to impose self-consistency by replacing the 
explicit expression for the Green's function in Eq.~(\ref{G_1}) inside the 
self-energy expressions Eqs.~(\ref{Sigma_1})-(\ref{Sigma_c}). After equating 
the scalar coefficients of the Pauli matrices this replacement leads finally 
to the anisotropic Eliashberg equations:
\begin{equation}
Z(\bbk,i\on) =
   1 + \frac{T}{\on \NF} \sum_{\bbkk n'}
   \frac{ \onp Z(\bbkk,i\onp) }{ \Theta(\bbkk,i\onp) } 
   \lambda(\bbk,\bbkk,n-n'),
\label{renormalization}
\end{equation}
\begin{equation}
\chi(\bbk,i\on) \!=
  -\frac{T}{\NF} \sum_{\bbkk n'}
   \frac{ \ekk + \chi(\bbkk,i\onp) }{ \Theta(\bbkk,i\onp) } 
   \lambda(\bbk,\bbkk\!,n-n'),
\label{shift}
\end{equation}
\begin{eqnarray}
\hspace{-1cm}\phi(\bbk,i\on) &=&
   \frac{T}{\NF} \sum_{\bbkk n'}
   \frac{ \phi(\bbkk,i\onp) }{ \Theta(\bbkk,i\onp) } \nonumber \\
  &&\times \left[ \lambda(\bbk,\bbkk,n-n') -\NF V(\bbk-\bbkk) \right].
\label{orderparam}
\end{eqnarray}
In Eqs.~(\ref{renormalization})-(\ref{orderparam}) $\NF$ represents
the density of states per spin at the Fermi level, and $\lambda(\bbk,\bbkk,n 
- n')$ is an auxiliary function describing the anisotropic electron-phonon 
coupling and defined as follows:
\begin{equation}
\lambda(\bbk,\bbkk,n - n') = \int_{0}^{\infty} d\omega  
\frac{2\omega}{(\on - \onp)^2+\omega^2}\afkko,
\end{equation}
with $\afkko$ the Eliashberg electron-phonon spectral function:
\begin{equation} \label{a2F_kko_1}
\afkko = \NF \sum_{\nu} |\gkkl|^2 \delta(\omega-\omega_{\bbk-\bbkk,\nu}).
\end{equation}
The superconducting gap is defined in terms of the renormalization function and
the order parameter as:
\begin{equation} \label{delta}
\Delta(\bbk,i\on) = \frac{\phi(\bbk,i\on)}{Z(\bbk,i\on)}.
\end{equation}
From Eqs.~(\ref{orderparam}),(\ref{delta}) we see that the Eliashberg 
equations admit the trivial solution $\Delta(\bbk,i\on)=0$ at all 
temperatures. The highest temperature for which the Eliashberg equations admit 
nontrivial solutions $\Delta(\bbk,i\on)\ne 0$ defines the critical temperature 
$T_{\rm c}$.

\subsection{Standard approximations} \label{sec.approx}

After having presented a concise derivation of Eliashberg's equations in the 
previous Section, we now discuss technical aspects which need to be addressed 
in order to actually solve these equations.

Since the superconducting pairing occurs mainly within an energy window of 
width $\omega_{\rm ph}$ around the Fermi surface ($\omega_{\rm ph}$ being a 
characteristic phonon energy), it is standard practice to simplify the 
Eliashberg equations by restricting the description to electron bands near the 
Fermi energy.~\cite{allen_mitrovic,marsiglio_book,scalapino66,allen76} This 
simplification can be achieved in the formalism by introducing the identity 
$\int_{-\infty}^{\infty} d\epsilon' \delta(\ekk\!-\!\epsilon')\!=\!1$ on the 
right hand side in Eqs.~(\ref{renormalization})-(\ref{orderparam}). The rapid 
changes of $\Theta(\bbkk,\onp)$ and the numerator of Eq.~(\ref{shift}) with 
the energy $\epsilon'$ can be integrated analytically, while for the other 
quantities we can set $\epsilon'$ to the Fermi energy since the associated 
variations take place on a much larger energy 
scale.~\cite{scalapino66,allen76} Under this approximation the energy shift 
becomes $\chi(\bbk,i\on)=0$ and only two equations are left to solve, one for 
the renormalization function and one for the order parameter (or equivalently 
the superconducting gap):
\begin{equation} 
Z(\bbk,i\on) = 
   1 + \frac{\pi T}{\on} \sum_{\bbkk n'} W_{\bbkk}
   \frac{ \onp }{ R(\bbkk,i\onp) } 
   \lambda(\bbk,\bbkk,n\!-\!n'),  
\label{Znorm_surf}   
\end{equation}
\begin{eqnarray} 
Z(\bbk,i\on) \Delta(\bbk,i\on) &=&  
   \pi T \sum_{\bbkk n'} W_{\bbkk} 
   \frac{ \Delta(\bbkk,i\onp) }{ R(\bbkk,i\onp) } \nonumber \\   
  &&\hspace{-2cm}\times\left[ \lambda(\bbk,\bbkk,\!n-\!n')-\NF 
  V(\bbk-\bbkk)\right],
\label{Delta_surf}
\end{eqnarray}
where $R(\bbk,i\on)$ and $W_{\bbk}$ are given by:
\begin{equation}\label{eq.R}
R(\bbk,i\on) = \sqrt{\on^2+\Delta^2(\bbk,i\on)} \textrm{ and }
W_{\bbk} = \delta(\ek)/\NF.
\end{equation}
Equations~(\ref{Znorm_surf}) and (\ref{Delta_surf}) form a coupled nonlinear 
system and need to be solved self-consistently at each temperature~$T$. The 
approximations leading to Eqs.~(\ref{Znorm_surf}) and (\ref{Delta_surf}) imply 
that $Z(\bbk,i\on)$ and $\Delta(\bbk,i\on)$ are only meaningful for the 
momentum/band index $\bbk$ at or near the Fermi surface. Away from the Fermi 
surface the energy dependence of these quantities is weak and is 
neglected.~\cite{scalapino66,allen76} In addition 
Eqs.~(\ref{Znorm_surf})-(\ref{Delta_surf}) implicitly assume that the
electron density of states is approximately constant near the Fermi energy.
This simplification may break down for materials with narrow bands or critical 
points in proximity of the Fermi level.~\cite{allen_mitrovic,pickett82}

In order to solve Eqs.~(\ref{Znorm_surf}),(\ref{Delta_surf}) numerically it is 
necessary to truncate the sum over Matsubara frequencies. It is standard 
practice to restrict the sum to frequencies smaller than a given cutoff 
$\omega_{\rm c}$, with the cutoff of the order of 1~eV and typically set to 
4-10 times the largest phonon energy. In addition, it is convenient to 
introduce a dimensionless Coulomb interaction parameter $\mu^*_{\rm c}$ 
defined as the double Fermi surface (FS) average over $\bbk$ and $\bbkk$ of 
the term $V(\bbk-\bbkk)$ in Eq.~(\ref{Delta_surf}):
\begin{equation}\label{muc0}
\mu_{\rm c} = \NF \langle\langle V(\bbk-\bbkk) \rangle\rangle_{\rm FS}.
\end{equation}
By performing the energy integral analytically up to the cutoff frequency it 
can be shown that the $\NF V(\bbk-\bbkk)$ term in Eq.~(\ref{Delta_surf}) can 
be replaced by the Morel-Anderson pseudopotential $\mu^*_{\rm c}$ given 
by:~\cite{Anderson}
\begin{equation}
\mu^*_{\rm c} = \frac{\mu_{\rm c}}{1+\mu_{\rm c}\ln (\EF/\omega_{\rm c})}.
\label{muc}
\end{equation}
Following this replacement, $\mu^*_{\rm c}$ is used as a semi-empirical 
parameter in the subsequent numerical solution of the Eliashberg equations.
For a large class of superconductors Eqs.~(\ref{muc0})-(\ref{muc}) yields 
values of $\mu^*_{\rm c}$ in the range 0.1-0.2. However, it is clear by now 
that in several cases values of the Coulomb parameter outside of this range 
are necessary for explaining experimental 
data.~\cite{liu_Pb,bauer,giustino_wannier} In addition, the anisotropic nature 
of the Coulomb interaction cannot be neglected for an accurate description of 
the superconducting properties.~\cite{golubov,mazin,floris_mgb2,moon} These 
observations should make it clear that the simplification provided by  
Eqs.~(\ref{muc0})-(\ref{muc}) is not optimal, and a fully {\it ab-initio} 
approach to the solution of the Eliashberg equations is highly 
desirable.~\cite{allen_mitrovic,pickett82} A description of the 
electron-phonon and the electron-electron interactions on the same footing is 
achieved in the SCDFT approach.~\cite{scdft1,scdft2} While it is clear that 
the Eliashberg approach considered in this work should be extended in order to 
incorporate Coulomb effects from first-principles, this is beyond the scope of 
the present investigation. 

\subsection{Superconducting gap along the real energy axis}

In Eqs.~(\ref{G_1})-(\ref{eq.R}) the dynamical aspects of the superconducting 
pairing are described using the imaginary Matsubara frequencies $i\omega_n$.
The reason for this choice is that the resulting formulation is computationally
efficient since it only involves sums over a finite number of frequencies.
While the imaginary axis formulation is adequate for calculating the critical
temperature as described in Sec.~\ref{sec.theory}, the calculation of spectral
properties such as the quasiparticle density of states requires the knowledge
of the superconducting gap along the real frequency axis. 

It is in principle possible to calculate the superconducting gap along the real
axis, however this procedure involves the evaluation of many principal value 
integrals and hence is numerically demanding.~\cite{carbotte,holcomb} In this 
work we prefer instead to determine the solutions of the Eliashberg equations 
on the real axis by analytic continuation of our calculated solutions along 
the imaginary axis. The analytic continuation can be performed either by using 
Pad\'{e} approximants as in Refs.~\onlinecite{vidberg,leavens} or by means of 
an iterative procedure as in Ref.~\onlinecite{marsiglio}.

The continuation based on Pad\'{e} approximants involves a very light 
computational workload, however it is very sensitive to the numerical 
precision of the solutions on the imaginary axis.~\cite{vidberg,leavens,beach}
As a rule of thumb the analytic continuation based on Pad\'{e} approximants 
exhibits the correct gross structure of the superconducting gap on the real 
frequency axis, however fine spectral features are not always captured 
completely.

The iterative analytic continuation, on the other hand, is generally rather 
accurate but involves a high computational workload. In fact, as shown in 
Ref.~\onlinecite{marsiglio}, the iterative analytic continuation requires 
solving the following equations self-consistently:
\begin{eqnarray} 
 Z(\bbk,\omega) \!& \!=\! & \!
   1 \!+ \!i \frac{\pi T}{\omega} \!\sum_{\bbkk n'} W_{\bbkk} 
   \frac{ \onp }{ R(\bbkk,i\onp) } \lambda(\bbk,\bbkk,\omega\!-\!i\onp) 
   \nonumber \\
   & + & i \frac{\pi}{\omega} \int_{-\infty}^{\infty} d\omega' 
   \Gamma(\omega,\omega')  
   \sum_{\bbkk} W_{\bbkk}      
   \alpha^2F(\bbk,\bbkk,\omega')  \nonumber \\
   &  \times &
   \frac{ (\omega-\omega') Z(\bbkk,\omega-\omega')}
        {P(\bbkk,\omega-\omega')},   
\label{Znorm_real}   
\end{eqnarray}
\begin{eqnarray} 
 && \hspace{-0.5cm} Z(\bbk,\omega) \Delta(\bbk,\omega) =  
   \pi T \sum_{\bbkk n'} W_{\bbkk}
   \left[ \lambda(\bbk,\bbkk,\omega-i\onp)- \mu^*_{\rm c} \right]
    \nonumber \\   
&&  \hspace{-0.5cm}\times \frac{ \Delta(\bbkk,i\onp) }{ R(\bbkk,i\onp) } 
  + i \pi  \int_{-\infty}^{\infty} \!\!d\omega' \Gamma(\omega,\omega') 
   \sum_{\bbkk} W_{\bbkk}
   \alpha^2F(\bbk,\bbkk,\omega')  \nonumber \\
   && \hspace{2.5cm}\times       
   \frac{ Z(\bbkk,\omega-\omega') \Delta(\bbkk,\omega-\omega')}
        {P(\bbkk,\omega-\omega')},  
\label{Delta_real}
\end{eqnarray}
where the following quantities have been introduced:
\begin{eqnarray} 
&& P(\bbk,\omega) =  
   \sqrt{Z^2(\bbk,\omega) \left[\omega^2 + \Delta^2(\bbk,\omega) 
   \right]},\label{eq.P}   \\
&& \Gamma(\omega,\omega') = \frac{1}{2} \left(
   \tanh{\frac{\omega-\omega'}{2T} } 
   + \coth{ \frac{\omega'}{2T} } \right),  \\
&& \lambda(\bbk,\bbkk,\omega-i\on) = 
   - \int_{-\infty}^{\infty} \!\!\!d\omega' 
   \frac{\alpha^2F(\bbk,\bbkk,\omega')}{\omega-i\on-\omega'},   \\
&& \alpha^2F(\bbk,\bbkk,-\omega)=-\alpha^2F(\bbk,\bbkk,\omega).
\end{eqnarray}
In the case where the square-root on the right hand side of Eq.~(\ref{eq.P}) 
is complex, the root with positive imaginary part is chosen. 

Once determined the mass renormalization function $Z(\bbk,\omega)$ and the superconducting 
gap $\Delta(\bbk,\omega)$ on the real frequency axis, one can examine the poles of the diagonal 
component of the single-particle Green's function:~\cite{marsiglio_book} 
\begin{eqnarray} \label{G11}
G_{11}(\bbk,\omega) =  
  \frac{ \omega Z(\bbk,\omega) + \ek }
       { [\omega Z(\bbk,\omega)]^2 - \ek^2 - [ Z(\bbk,\omega) 
       \Delta(\bbk,\omega)]^2},
\end{eqnarray} 
in order to obtain the quasiparticle energy $E_{\bbk}$:
\begin{eqnarray} 
E_{\bbk}^2 = 
  \left[ \frac{\ek}{Z(\bbk,E_{\bbk})} \right]^2 +     
  \Delta^2(\bbk,E_{\bbk}).        
\end{eqnarray}
At the Fermi level $\ek=0$ and the quasiparticle shift is
$E_{\bbk}=\textrm{Re}\Delta(\bbk,E_\bbk)$. As a result the 
leading edge $\Delta_\bbk$ of the superconducting gap for the 
combined band/momentum index $\bbk$ at the Fermi surface is given by:
\begin{equation}
\text{Re}[\Delta(\bbk,\Delta_\bbk)]=\Delta_\bbk. 
\end{equation}

\subsection{Isotropic approximation}
\label{sec.isotropic}

For conventional bulk metals or superconductors where the Fermi surface 
anisotropy is either weak or smeared out by impurities, it is possible to
resort to a simplified isotropic formulation of the Eliashberg equations. 
Such formulation is obtained from Eqs.~(\ref{Znorm_surf}),(\ref{Delta_surf})
by averaging $\bbk$ over the Fermi surface. We obtain:
\begin{equation} 
Z(i\on) = 
   1 + \frac{\pi T}{\on} \sum_{n'} 
   \frac{ \onp }{ R(i\onp) } \lambda(n\!-\!n'),  
\label{Znorm_iso}   
\end{equation}
\begin{eqnarray} 
Z(i\on) \Delta(i\on) =  
   \pi T \sum_{n'} 
   \frac{ \Delta(i\onp) }{ R(i\onp) } 
  \left[ \lambda(n\!-\!n')-\mu_{\rm c}^*\right],
\label{Delta_iso}
\end{eqnarray} 
where $R(i\on)$ and $\lambda(n-n')$ are given by:

\begin{equation}
R(i\on) = \sqrt{\on^2+\Delta^2(i\on)},
\end{equation}
\begin{equation}
\lambda(n - n') = \int_{0}^{\infty} d\omega  
\frac{2\omega\afo}{(\on - \onp)^2+\omega^2},
\end{equation}
and $\afo$ is the isotropic Eliashberg spectral function:
\begin{equation} \label{a2F}
\afo = \sum_{\bbk,\bbkk} W_{\bbk} W_{\bbkk} \afkko.
\end{equation}
The isotropic Eliashberg equations on the real axis can be obtained similarly 
by starting from Eqs.~(\ref{Znorm_real}),(\ref{Delta_real}). From the 
isotropic superconducting gap on the real axis we can obtain the normalized 
quasiparticle density of states in the superconducting state $N_S(\omega)$:
\begin{equation} \label{eq.qdos}
\frac{N_S(\omega)}{N_F}=
\text{Re} \left[ \frac{\omega}{\sqrt{\omega^2-\Delta^2(\omega)}} 
\right].
\end{equation}

\section{COMPUTATIONAL METHODOLOGY}\label{section.methodology}

\subsection{Electron-phonon Wannier interpolation}

We now describe the combination of the anisotropic Eliashberg formalism of 
Sec.~\ref{sec.theory} with the electron-phonon Wannier interpolation of 
Ref.~\onlinecite{giustino_wannier}. The numerical solution of 
Eqs.~(\ref{Znorm_surf}),(\ref{Delta_surf}) and 
Eqs.~(\ref{Znorm_real}),(\ref{Delta_real}) requires an extremely careful 
description of the electron-phonon scattering processes, especially in 
proximity of the Fermi surface. This requirement translates into the necessity 
of evaluating electronic eigenvalues $\ek$, phonon frequencies $\omega_{\bbq 
\nu}$, and electron-phonon matrix elements $g_{\bbk\bbkk\nu}$ for a very large 
set of electron and phonon wavevectors in the Brillouin zone, of the order of 
tens of thousands. 

While it is practically impossible to evaluate so many electron-phonon matrix 
elements directly using standard density-functional calculations, it is 
possible to perform an optimal {\it ab-initio} interpolation of the matrix 
elements by exploiting localization in real space. The key idea is to first 
evaluate a small number of electron-phonon matrix elements in the 
maximally-localized Wannier representation, and then perform a generalized
Fourier interpolation into the momentum space, i.e. into the Bloch 
representation. The relation between the matrix elements in the Wannier 
representation $g_{{\bf R}{\bf R}}$ and those in the Bloch representation 
$g_{\bbk\bbkk}$ is:
\begin{equation} \label{eq.wannier}
  g_{\bbk\bbkk} = \frac{1}{N} \sum_{{\bf R},{\bf R}'} {\rm e}^{i (\bbk \cdot 
                              {\bf R}+\bbq\cdot{\bf R}')} \,
         \, U_{\bbkk} \, g_{{\bf R}{\bf R}'}  \, U^\dagger_\bbk u_{\bbq},
\end{equation}
where $N$ is the size of the discrete Brillouin-zone mesh, $U_\bbk$ is a 
band-mixing matrix which maps electron Bloch bands into Wannier functions, 
$u_{\bbq}$ is a branch-mixing matrix which maps phonon branches into 
individual atomic displacements, $\bbq\!=\!\bbkk\!-\bbk$, and ${\bf R},{\bf 
R}'$ are vectors of the direct lattice. In Eq.~(\ref{eq.wannier}) the band and 
branch indices are absorbed in $\bbk,\bbkk$ and in ${\bf R},{\bf R}'$. More 
detailed expressions for implementation purposes can be found in 
Refs.~\onlinecite{giustino_wannier,EPW}. Once obtained the matrix elements in 
the Wannier representation, the evaluation of Eq.~(\ref{eq.wannier}) for any 
pairs of initial and final electron wavevectors is inexpensive since it 
involves only very small matrix multiplications.

The matrix elements in the Wannier representation are computed by first 
calculating the corresponding elements in the Bloch representation on a coarse 
Brillouin zone mesh using density-functional perturbation 
theory~\cite{baroni2001} and then transforming into the maximally localized 
Wannier representation~\cite{marzari1997,souza2001} using the inverse relation 
of Eq.~(\ref{eq.wannier}). All our density-functional and density-functional 
perturbation theory calculations are performed using the {\tt Quantum 
ESPRESSO} package,~\cite{QE} and maximally-localized Wannier functions are 
determined using the {\tt Wannier90} program.~\cite{WANNIER} The subsequent 
electron-phonon interpolation is performed using the {\tt EPW} 
program,~\cite{EPW} which extracts and processes information from both {\tt 
Quantum ESPRESSO} and {\tt Wannier90}. Further details on the notion of 
Wannier interpolation and its use in the study of electron-phonon interactions 
can be found in Ref.~\onlinecite{marzari2012} and 
Refs.~\onlinecite{giustino_wannier,EPW}, respectively.

Even when using electron-phonon Wannier interpolation the computational 
workload can become quite substantial when one evaluates hundreds of thousands 
of matrix elements. In order to reduce the computational load we exploit the 
crystal symmetries and only evaluate the gap function $\Delta(\bbk,i\omega_n)$ 
and the renormalization function $Z(\bbk,i\omega_n)$ in the irreducible wedge 
of the Brillouin zone. On the other hand, the sums over $\bbkk$ in 
Eqs.~(\ref{Znorm_surf}),(\ref{Delta_surf}) and 
Eqs.~(\ref{Znorm_real}),(\ref{Delta_real}) are performed over the entire 
Brillouin zone. The meshes of wavevectors $\bbk$ and 
$\bbq\!=\!\bbkk\!-\bbk$ are chosen to be uniform and commensurate, in such a 
way that the grid of electron wavevectors in the final state $\bbkk$ maps 
into the grid of the initial wavevectors $\bbk$. Since the contributions to 
the superconducting gap arising from electronic states away from the Fermi 
energy are essentially negligible, the matrix elements of 
Eq.~(\ref{eq.wannier}) are evaluated only for electronic states such that 
$\epsilon_\bbk$ and $\epsilon_\bbkk$ are near the Fermi energy. Numerical 
convergence can be achieved typically by restricting the sums in 
Eqs.~(\ref{Znorm_surf}),(\ref{Delta_surf}) and 
Eqs.~(\ref{Znorm_real}),(\ref{Delta_real}) to an energy window around the 
Fermi level of width corresponding to 4-10 times the characteristic phonon 
frequency.

\subsection{Self-consistent solution of the nonlinear system and analytic 
continuation} \label{subsection.cont}

In order to solve iteratively the Eliashberg equations on the imaginary axis 
Eqs.~(\ref{Znorm_surf}),(\ref{Delta_surf}), we start from an initial guess 
$\Delta_0(i\omega_n)$ for the superconducting gap. The starting guess 
$\Delta_0(i\omega_n)$ is chosen to be a step function vanishing for $i 
\omega_n>2 \omega_{\rm ph}^{\rm max}$, $\omega_{\rm ph}^{\rm max}$ being the 
largest phonon energy in the system. The magnitude of $\Delta_0(i\omega_n)$ is 
estimated from the BCS formula~\cite{bcs} at zero temperature 
$2\Delta_0(i\omega_n)/T_{\rm c}=3.52$, with $T_{\rm c}$ given by Allen-Dynes 
equation.~\cite{allen-dynes}  

Our experience shows that the convergence of the iterative self-consistent 
solution is significantly improved by using the Broyden mixing scheme commonly 
employed in standard density-functional calculations.~\cite{broyden,johnson} 
For the test cases considered in Sec.~\ref{section.applications} below we find 
that around 15-20 iterations are sufficient to achieve convergence whenever
$T\lesssim 0.8T_{\rm c}$. The number of iterations increases to 40-60 for 
temperatures between 0.8-0.95$T_{\rm c}$, and may exceed 100 for 
$T \gtrsim 0.95 T_{\rm c}$. In order to accelerate the convergence we use the 
gap functions calculated at a given temperature as seeds for the iterations at 
the next temperature. An alternative strategy for solving the equations when
$T\! \simeq\! T_{\rm c}$ would be to use the linearized form of the Eliashberg 
equations and determine the critical temperature by solving an eigenvalue 
problem,\cite{allen_mitrovic,choi_mgb2} however we did not explore this 
possibility. 

In order to determine the superconducting gap along the real energy axis we
consider two possibilities. The first possibility is to perform an approximate
analytic continuation using Pad\'{e} functions.~\cite{vidberg,leavens,beach}
This procedure works well if the Pad\'{e} functions are constructed using
the Matsubara frequencies on the imaginary axis. The second possibility 
consists of performing the exact analytic continuation of the imaginary 
solution to the real energy axis as described in Sec.~\ref{subsection.cont}.
Since this latter approach is computationally very demanding, we speed up the
convergence of the iterative analytic continuation by using the approximate 
Pad\'{e} continuation as an initial guess.

\section{APPLICATIONS}\label{section.applications}

In order to validate the computational methodology developed within {\tt EPW}, 
we investigate two prototypical systems: the nearly-isotropic lead (Pb)
superconductor, and the anisotropic magnesium diboride (MgB$_2$) 
superconductor. 

\subsection{Computational details}

The calculations are performed within the local density approximation (LDA) 
to density-functional theory~\cite{lda1,lda2} using {\tt Quantum 
ESPRESSO}.~\cite{QE} The valence electronic wavefunctions are expanded in 
planewaves basis sets with kinetic energy cutoff of 80 Ry and 60 Ry for Pb and 
MgB$_2$, respectively. The core-valence interaction is taken into account by 
using norm-conserving pseudopotentials.~\cite{nc1,nc2} For Pb we consider four 
valence electrons and a scalar-relativistic pseudopotential. In order to 
facilitate the comparison with previous theoretical studies we use the LDA 
theoretical lattice parameters for Pb ($a=4.778$~\AA) and the experimental 
lattice parameters for MgB$_2$ ($a=3.083$ \AA\ and $c/a=1.142$). The charge 
density is computed using $\Gamma$-centered Brillouin-zone mesh with $16^3$ 
and $24^3$ $\bbk$-points for Pb and MgB$_2$, respectively, and a 
Methfessel-Paxton smearing~\cite{mp} of 0.02~Ry. The dynamical matrices and 
the linear variation of the self-consistent potential are calculated within 
density-functional perturbation theory~\cite{baroni2001} on the irreducible 
set of a regular $8^3$ (Pb) and $6^3$ (MgB$_2$) $\bbq$-point meshes. The 
electronic wavefunctions required for the Wannier interpolation within {\tt 
EPW} are calculated on uniform and $\Gamma$-centered $\bbk$-point meshes of 
sizes $8^3$ and $6^3$ for Pb and MgB$_2$.

In the case of Pb four Wannier functions are used to describe the electronic 
structure near the Fermi level. These states are $sp^3$-like functions 
localized along each one of the Pb-Pb bonds, with a spatial spread of 
2.40~\AA. In the case of MgB$_2$ we consider five Wannier functions in order 
to describe the band structure around the Fermi level. Two functions are 
$p_z$-like states and are associated with the B atoms, and three functions are 
$\sigma$-like states localized in the middle of B-B bonds. The spatial spread 
of the MLWFs in MgB$_2$ are 2.02~\AA\ ($p_z$) and 1.16~\AA\ ($\sigma$). 

In order to solve the Eliashberg equations we evaluate electron energies,
phonon frequencies, and electron-phonon matrix elements on fine grids
using the method of Ref.~\onlinecite{giustino_wannier}. The fine grids
contain $(40^3,40^3)$ $\bbk$- and $\bbq$-points for Pb (random grids),
and $(60^3,30^3)$ points for MgB$_2$ (uniform $\Gamma$-centered grids).
Such an extremely fine sampling of the Brillouin zone is found to be crucial 
for the convergence of the superconducting energy gap in the fully anisotropic 
case. The frequency cutoff $\omega_{\rm c}$ in 
Eqs.~(\ref{Znorm_surf}),(\ref{Delta_surf}) and 
Eqs.~(\ref{Znorm_iso}),(\ref{Delta_iso}) is set to ten times the maximum 
phonon frequency of the system: $\omega_{\rm c}=10\omega_{\rm ph}^{\rm max}$. 
The calculations are performed using smearing parameters in the Dirac delta 
functions corresponding to 100~meV and 50~meV for electrons and phonons, 
respectively.

\subsection{Lead}

\begin{figure}[t]
\centerline{\includegraphics[width=85mm]{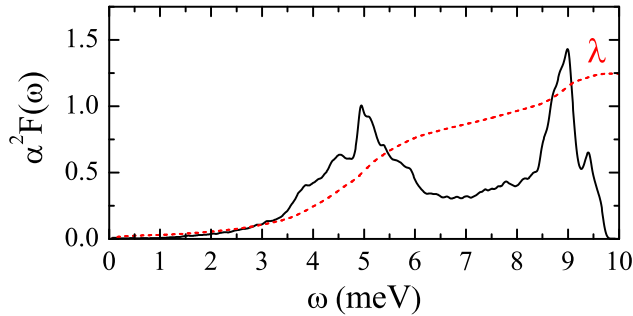}}	
\caption{ (Color online) Calculated isotropic Eliashberg spectral function 
$\alpha^2F$ (black solid line) of Pb, and cumulative contribution
to the electron-phonon coupling strength $\lambda$ (red dashed line). 
The top of the red dashed curve corresponds to $\lambda$=1.24. }

\label{lambda_Pb}
\end{figure}

Bulk lead is the best known example of a strong-coupling superconductor, 
exhibiting a superconducting transition temperature 
$T_{\rm c}=7.2$~K.~\cite{Pb_exp} Although Pb is known to be a two-band 
superconductor, the superconducting gap function is only very weakly 
anisotropic,~\cite{blackford,lykken,tomlinson_Pb,floris_Pb} therefore for the 
sake of testing our method we use the isotropic approximation to the 
Migdal-Eliashberg formalism described in Sec.~\ref{sec.isotropic}.

\begin{figure}[t]
\centerline{\includegraphics[width=85mm]{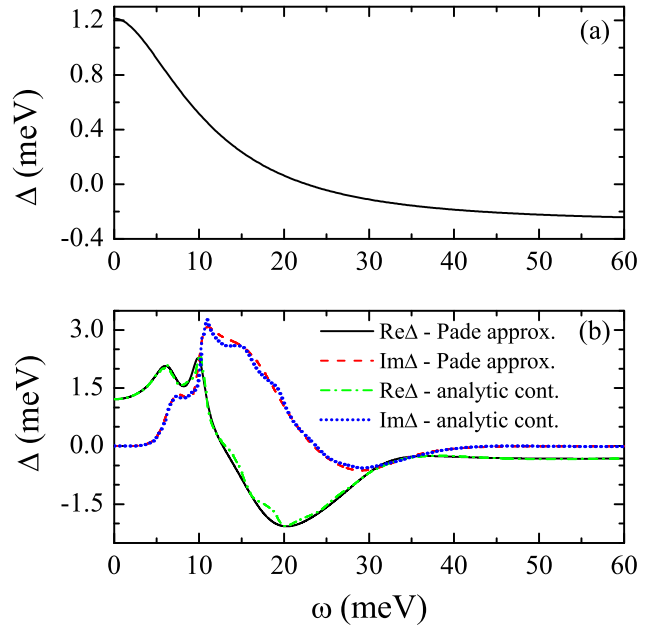}}	
\caption{ (Color online) Calculated energy-dependent superconducting gap of Pb 
at $T$=0.3~K. The gap is obtained by solving the isotropic Eliashberg equations
with $\mu_{\rm c}^*=0.1$. (a) Superconducting gap along the imaginary energy 
axis (black solid line). (b) Superconducting gap along the real energy axis.
We show both the solutions obtained from the approximate analytic continuation
using  Pad\'{e} functions (black solid line and red dashed line), and the 
solutions obtained using the iterative analytic continuation (green 
dash-dotted line and blue dotted line). }
\label{gap_Pb}
\end{figure} 

Figure~\ref{lambda_Pb} shows the calculated Eliashberg spectral function 
$\alpha^2F(\omega)$ and the corresponding electron-phonon coupling parameter 
$\lambda$. We find an overall good agreement with experimental 
results,~\cite{Pb_exp} although we observe a small ($\simeq$0.5~meV) 
but non-negligible blueshift of the two peaks in the Eliashberg function. 
This blueshift is well understood now and arises from the overestimation of 
the phonon frequencies in the absence of spin-orbit coupling in our 
calculation.~\cite{dalcorso,heid} Our calculated electron-phonon coupling
$\lambda$=1.24 lies in between the values reported in previous theoretical 
studies,~\cite{floris_Pb,tomlinson_Pb,liu_Pb} although it is somewhat smaller 
than the value 1.55 obtained from tunneling measurements~\cite{Pb_exp}
owing to the neglect of spin-orbit coupling.

Figure~\ref{gap_Pb} shows the solutions of the isotropic Eliashberg equations 
Eqs.~(\ref{Znorm_iso}),(\ref{Delta_iso}) for $\mu_{\rm c}^*=0.1$ and $T$=0.3~K.
Along the imaginary axis the superconducting gap function is purely real
and displays a frequency dependence similar to standard plasmon-pole models
[Fig.~\ref{gap_Pb}(a)]. The continuation of the calculated superconducting gap 
to the real energy axis is shown in Fig.~\ref{gap_Pb}(b). We see that the 
approximate analytic continuation using Pad\'{e} functions and the exact 
iterative analytic continuation yield very similar results. As expected the 
approximate Pad\'{e} continuation misses some of the fine features which are 
instead observed in the exact iterative continuation. Our calculated 
superconducting gap is in very good agreement with solutions of the Eliashberg 
equations obtained directly on the real energy axis.~\cite{tomlinson_Pb}
In Fig.~\ref{gap_Pb}(b) we see that a two-peak structure emerges both in the 
real part and the imaginary part of $\Delta(\omega)$. These two peaks occur on 
the scale of the phonon energies and correlate with those observed in the 
Eliashberg spectral function of Fig.~\ref{lambda_Pb}.~\cite{scalapino} A 
detailed analysis shows that the peaks in the real part of the gap function 
are blueshifted by approximately $\Delta_0=\Delta(\omega\!=\!0)$ with respect 
to the corresponding peaks in $\alpha^2F(\omega)$.     

\begin{figure}[t]
\centerline{\includegraphics[width=85mm]{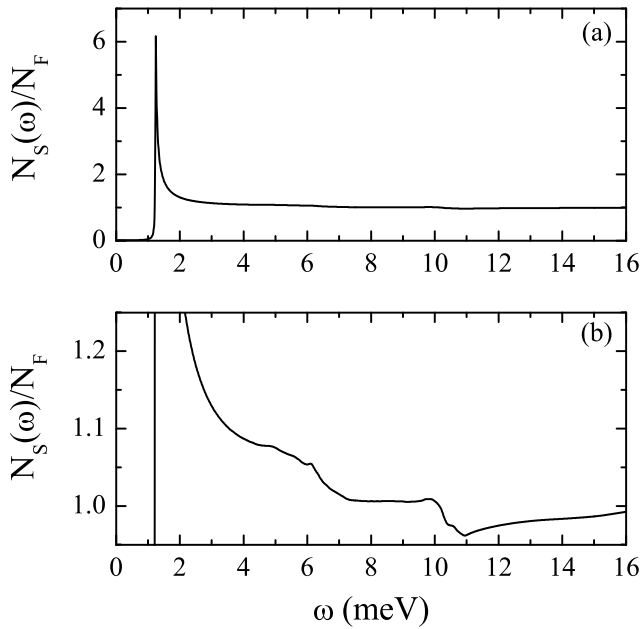}}	
\caption{ (Color online) (a) Calculated quasiparticle density of states of Pb
at $T$=0.3~K (black solid lines). The superconducting gap is obtained from 
Fig.~\ref{gap_Pb}. (b) Same quantity as in (a), magnified in order to show the 
structure which is used in tunneling experiments for extracting the Eliashberg
spectral function.}
\label{qdos_Pb}
\end{figure}

Figure~\ref{qdos_Pb}(a) shows the normalized quasiparticle density of  
unoccupied states obtained from Eq.~(\ref{eq.qdos}) using the gap function of 
Fig.~\ref{gap_Pb}. As expected the strong van Hove singularity marks the 
leading edge $\Delta_0$ of the superconducting gap. The fine structure of the 
density of states around the van Hove singularity [Fig.~\ref{qdos_Pb}(b)] is 
the direct signature of the electron-phonon physics and is precisely the basis 
for direct measurements of the Eliashberg function using tunneling 
spectroscopy.

\begin{figure}[t]
\centerline{\includegraphics[width=80mm]{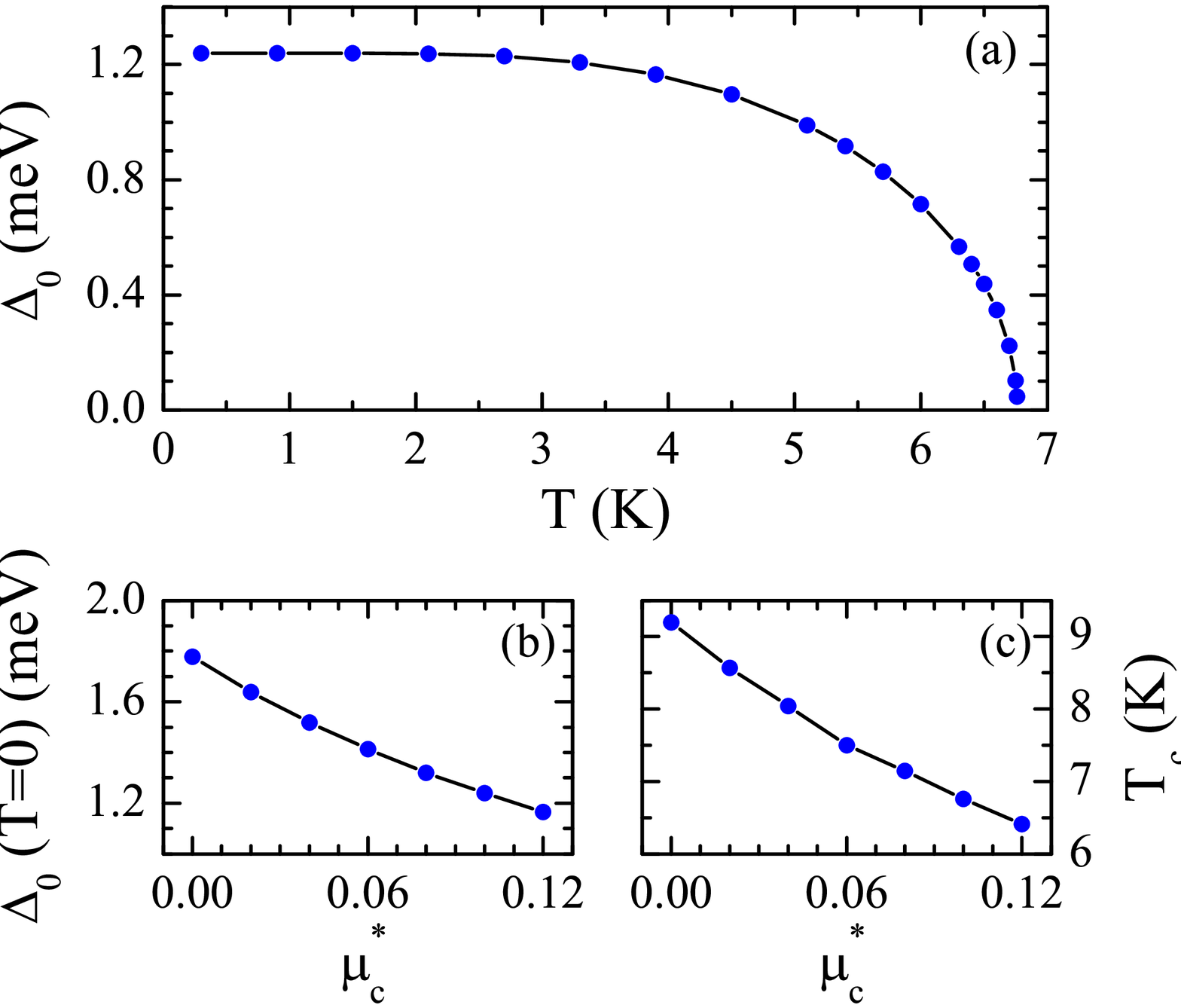}}	
\caption{ (Color online) (a) Calculated superconducting gap of Pb at the Fermi 
level as a function of temperature (disks). The Coulomb parameter is set to 
$\mu_{\rm c}^*=0.1$. (b) Calculated superconducting gap of Pb at the Fermi 
level for $T$=0~K as a function of the Coulomb parameter $\mu_{\rm c}^*$ (disks).
(c) Calculated critical temperature as a function of the Coulomb parameter 
$\mu_{\rm c}^*$ (disks). In all panels the solid lines are guides to the eye.}
\label{gap0_Pb}
\end{figure}

Figure~\ref{gap0_Pb} shows the superconducting gap function at the Fermi
level as a function of temperature, calculated for $\mu_{\rm c}^*=0.1$.
The leading edge of the gap at $T$=0~K is found to be $\Delta_0$=1.24~meV,
in good agreement with tunneling measurements yielding 1.33~meV.~\cite{Pb_exp}
The superconducting $T_{\rm c}$ is identified as the temperature at which
the gap vanishes. From Fig.~\ref{gap0_Pb}(a) we find $T_{\rm c}$=6.8~K,
in very good agreement with previous theoretical 
studies,~\cite{scdft2,tomlinson_Pb} and only slightly lower than the 
experimental datum of 7.2~K. For completeness in Fig.~\ref{gap0_Pb}(b) and (c) 
we also explore the sensitivity of the calculated gap and critical temperature 
to the choice of the Coulomb parameter $\mu_{\rm c}^*$. As expected, a 
reduction of the effective Coulomb interaction results in an increase of both 
$\Delta_0$ and $T_{\rm c}$.

\subsection{Magnesium diboride}

Within the class of phonon-mediated superconductors MgB$_2$ holds the
record of the highest critical temperature, with $T_{\rm c}$=39~K.\cite{MgB2_exp} After a decade of intense experimental and theoretical investigations since its discovery, it is now understood that MgB$_2$ is an anisotropic two-gap electron-phonon superconductor.~\cite{kortus,liu_mgb2,giubileo,choi_mgb2,floris_mgb2}
The anisotropy of the superconducting gap is a consequence of the multi-sheet 
Fermi surface of MgB$_2$, consisting of two hole-like coaxial cylinders 
arising from the $\sigma$ bonding bands, and two hole-like tubular networks 
arising from the $\pi$ bonding and antibonding bands (see for example Fig.~3 
of Ref.~\onlinecite{kortus}).

\begin{figure}[t]
\centerline{\includegraphics[width=85mm]{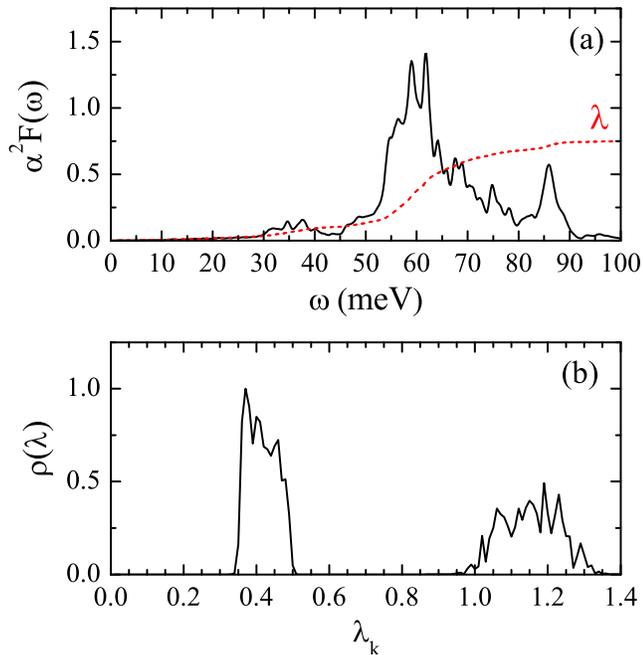}}
\caption{ (Color online) (a) Calculated isotropic Eliashberg spectral function
$\alpha^2F$ of MgB$_2$ (black solid line) and cumulative contribution 
to the electron-phonon coupling strength $\lambda$ (red dashed line). 
(b) Distribution of the electron-phonon coupling strenght $\lambda_\bbk$
for MgB$_2$ (black solid line). }
\label{lambda_aniso}
\end{figure}

Figure~\ref{lambda_aniso}(a) shows our calculated isotropic Eliashberg 
spectral function $\alpha^2F(\omega)$ and electron-phonon coupling strength 
$\lambda$ for MgB$_2$. $\alpha^2F(\omega)$ displays a large dominant peak 
centered around 60~meV and a second weaker peak centered around 86~meV. The 
corresponding isotropic electron-phonon coupling strength is $\lambda$=0.748. 
In order to quantify the different contributions to the coupling strength
associated with the $\sigma$ sheets and from the $\pi$ sheets of the Fermi 
surface we evaluate a band- and wavevector- dependent electron-phonon coupling 
strength defined by $\lambda_\bbk = \sum_{\bbkk} W_{\bbkk} 
\lambda(\bbk,\bbkk,n\!=\!0)$. Figure~\ref{lambda_aniso}(b) shows that the 
calculated $\lambda_\bbk$ cluster into two separate ranges. The lower range 
$\lambda_\bbk$=0.35-0.50 corresponds to the coupling of the $\pi$ Fermi 
surface sheets, and the higher range 0.95-1.35 corresponds to the coupling of 
the $\sigma$ sheets. The wider range of $\lambda_\bbk$ in the $\sigma$ sheets 
reflects a more pronounced anisotropy with respect to the $\pi$ sheets.
The structure of $\alpha^2F(\omega)$ and the calculated coupling strength are
in good agreement with previous 
calculations.~\cite{bohnen,choi_mgb2,floris_mgb2,eiguren,calandra}
In particular our results are in very good agreement with those reported in
Ref.~\onlinecite{eiguren} where a related interpolation scheme was used.
For completeness we show in Table~\ref{table1} the sensitivity of the 
calculated average coupling strength $\lambda$ on the underlying 
Brillouin-zone grids, and we compare with previous first-principles 
calculations.

\begin{table}[b]
\begin{center}
\begin{tabular}[t]{llrcrcl}\hline\hline \\
Reference & \hspace{0.2cm} & $\bbk$-mesh & \hspace{0.5cm} & $\bbq$-mesh & 
\hspace{0.2cm} & $\lambda$ \\\\\hline
Bohnen {\it et al.} (Ref.~\onlinecite{bohnen}) & & $36^3$ & & $6^3$ & & 0.73 
\\ 
Choi {\it et al.} (Ref.~\onlinecite{choi_mgb2}) & & $12\!\times\!18^2$ & & 
$12\!\times 
18^2$ & & 0.73 \\ 
Floris {\it et al.} (Ref.~\onlinecite{floris_mgb2}) & & $24^3$ & & $20^3$ & & 
0.71 
\\
Eiguren {\it et al.} (Ref.~\onlinecite{eiguren}) & & $40^3$ & & $40^3$ & & 
0.776 \\ 
Calandra {\it et al.} (Ref.~\onlinecite{calandra}) & & $80^3$ & & $20^3$ & & 
0.741 
\\
\\
This work & & $40^3$ & & $20^3 \,(40^3)$ & & 0.735 \\
& & $80^3$ & & $20^3 \,(40^3)$ & & 0.739 \\
& & $60^3$ & & $30^3 \,(60^3)$ & & 0.748 \\
& & $30^3$ & & $30^3$ & & 0.782 \\
& & $50^3$ & & $50^3$ & & 0.744 \\
& & 64000 & & 8000 & & 0.757 \\
& & 216000 & & 27000 & & 0.726 \\
\hline \hline
\end{tabular}
\caption{\small Electron-phonon coupling strength $\lambda$ of MgB$_2$ 
calculated using various meshes of $\bbk$- and $\bbq$-points in the Brillouin 
zone. The numbers in the brackets correspond to a second choice of 
$\bbq$-mesh while keeping the $\bbk$-mesh unchanged. The two bottom rows 
correspond to uniform random distributions of $\bbk$- and $\bbq$-points.}
\label{table1}
\end{center}
\end{table}

\begin{figure}[t]
\centerline{\includegraphics[width=85mm]{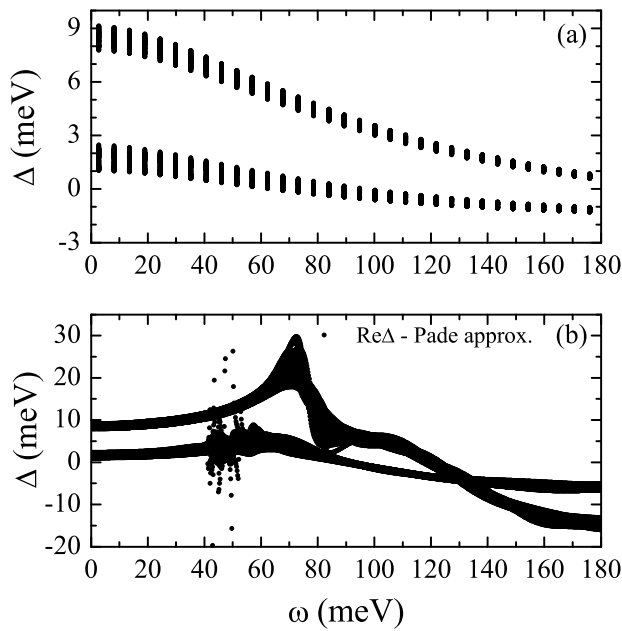}}	
\caption{
(Color online)
Calculated energy-dependent superconducting gap of MgB$_2$ at $T$=10~K.
The gap is obtained by solving the fully anisotropic Eliashberg equations
with $\mu_{\rm c}^*=0.16$. (a) Superconducting gap along the imaginary energy 
axis (black dots). (b) Superconducting gap along the real energy axis, 
obtained from the approximate analytic continuation using  Pad\'{e} functions 
(black dots). Only a representative sample of $10^5$ data points out of entire 
set of calculated gaps ($10^7$ points) is shown for clarity.} 
\label{gap_MgB2}
\end{figure}

Figure~\ref{gap_MgB2}(a) shows the anisotropic superconducting gap function
$\Delta(\bbk,\omega)$ of MgB$_2$ at $T$=10~K calculated along the imaginary 
axis using the anisotropic Eliashberg equations 
Eqs.~(\ref{Znorm_surf}),(\ref{Delta_surf}) 
and $\mu^*_{\rm c}$=0.16. Figure~\ref{gap_MgB2}(b) shows the real part of the 
superconducting gap function along the real energy axis, as obtained from the 
imaginary axis solutions of Fig~\ref{gap_MgB2}(a) via the approximate analytic 
continuation using Pad\'{e} functions. The two-gap nature of MgB$_2$ emerges 
in a completely natural way from our implementation. Indeed for each energy 
two distinct sets of superconducting gaps can be identified and associated 
with the $\sigma$ and the $\pi$ sheets of the Fermi surface. The two gaps are 
both anisotropic, and the corresponding Fermi-surface averages are 
$\Delta_{\pi}=1.8$~meV and $\Delta_{\sigma}=8.5$~meV, respectively. For 
comparison the experimental values for the gaps lie in the range 2.3-2.8~meV 
for $\pi$-band, and 7.0-7.1~meV for 
$\sigma$-band.~\cite{iavarone,szabo,gonnelli} As in the case of Pb, the 
structure that can be observed both in the $\sigma$ and $\pi$ superconducting 
gaps reflect the peaks occuring in the Eliashberg spectral function of 
Fig.~\ref{lambda_aniso}.

\begin{figure}[t]
\centerline{\includegraphics[width=85mm]{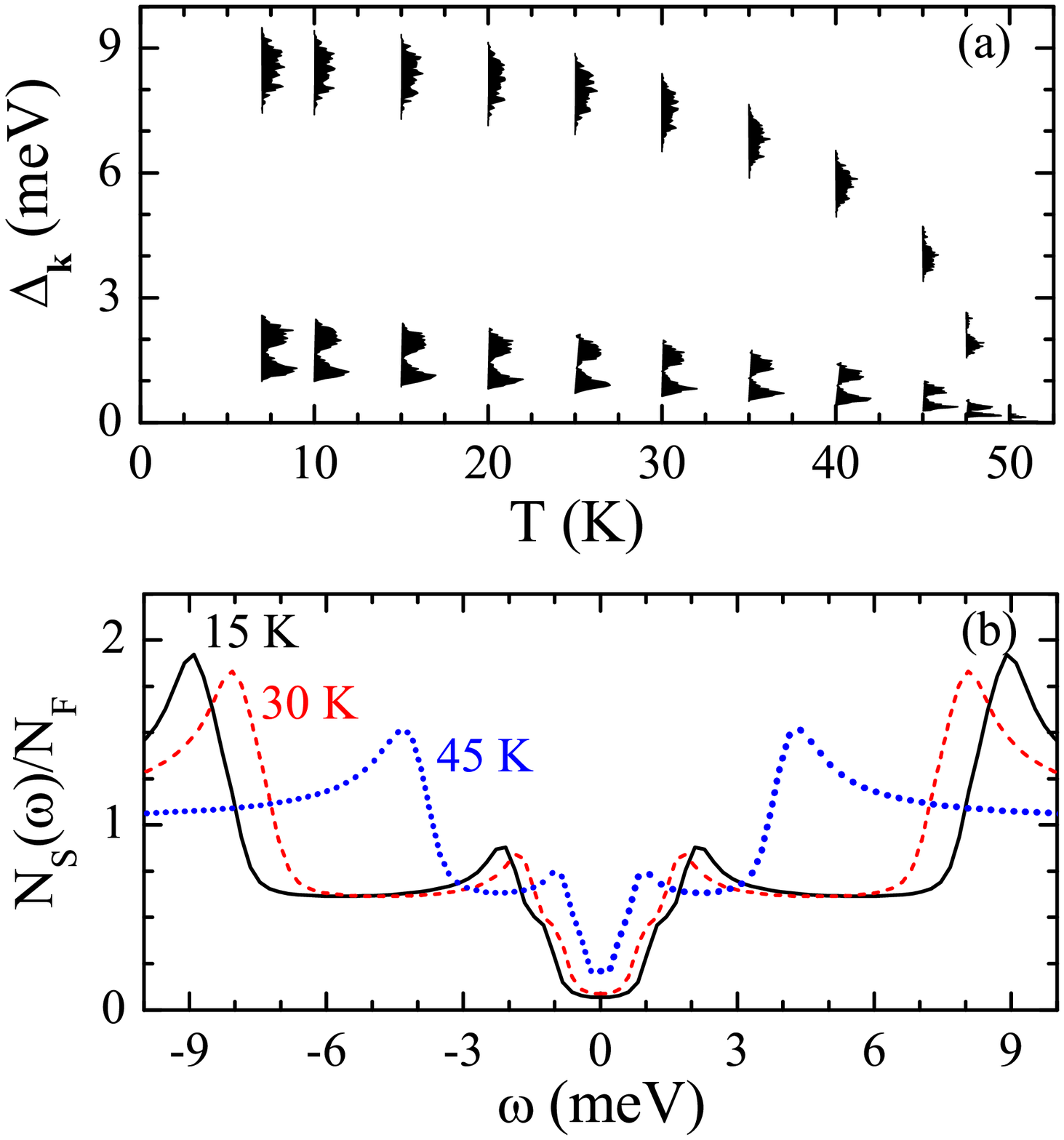}}	
\caption{ (Color online) (a) Calculated anisotropic superconducting gaps of 
MgB$_2$ on the Fermi surface as a function of temperature. The Coulomb 
potential is set to $\mu^*_{\rm c}=0.16$. (b) Corresponding quasiparticle 
density of states for a few representative temperatures (15~K black solid 
line, 30~K red dashed line, 45~K blue dash-doted line). } 
\label{gap0_qdos}
\end{figure}

Figure~\ref{gap0_qdos}(a) shows the calculated leading edges $\Delta_\bbk$ of 
the superconducting gaps as a function of temperature. Both the $\pi$ and 
$\sigma$ gaps vanish at the critical temperature $T_{\rm c}$=50~K. The 
corresponding quasiparticle density of states, presented in 
Fig.~\ref{gap0_qdos}(b), clearly shows the two-gaps structure of MgB$_2$,
in agreement with experiment.~\cite{giubileo}

Our calculated critical temperature is larger than the experimentally measured 
$T_{\rm c}$ of 39~K,~\cite{MgB2_exp} however it is in very good agreement with 
previous first-principles calculations based on the ME 
formalism~\cite{choi_mgb2} or the SCDFT formalism.~\cite{floris_mgb2} 
At this time it is still unclear whether the overestimation of the experimental
critical temperature is due to possible anharmonic effects~\cite{choi_mgb2}
or to the use of an isotropic Coulomb parameter.~\cite{floris_mgb2}
In fact, using the anisotropic Eliashberg formalism and a Coulomb parameter 
$\mu^*_{\rm c}=0.12$, the authors of Ref.~\onlinecite{choi_mgb2} find that the 
calculated $T_{\rm c}$ decreases from 55~K to 39~K if phonon anharmonicity is 
taken into account. On the other hand, using the SCDFT formalism the authors 
of Ref.~\onlinecite{floris_mgb2} calculate a critical temperature $T_{\rm 
c}$=22~K when using the complete wavevector-dependent superconducting gap.
When employing a band-averaged superconducting gap and various levels of 
approximations for the Coulomb interaction, the same authors find critical 
temperatures in the range 30-50~K.~\cite{floris_mgb2} A similar sensitivity of 
the calculated $T_{\rm c}$ to fine details of the calculations are reported in 
other studies based on a two-bands approximation to the ME 
formalism.~\cite{mazin,moon} The origin of the discrepancy between 
first-principles calculations of the critical temperature of MgB$_2$ and 
experiment clearly deserves further investigation, however this is beyond the 
scope of the present manuscript.

\begin{figure}[t]
\centerline{\includegraphics[width=85mm]{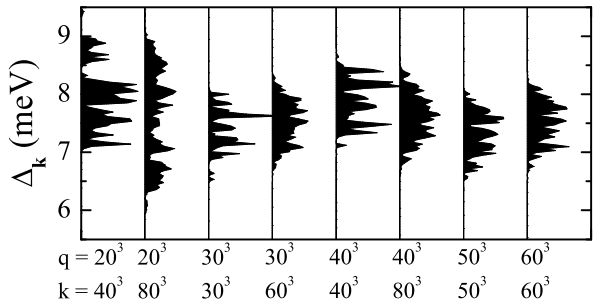}}	
\caption{ (Color online) Calculated energy distribution of the superconducting 
gaps on the $\sigma$ sheets of the Fermi surface of MgB$_2$ at $T$=30~K (black 
lines). The gaps are obatined for various $\bbk$- and $\bbq$-points meshes, as 
indicated by the labels on the horizontal axis. The Coulomb parameter is set 
to $\mu_{\rm c}^*=0.16$.}
\label{gap0_mesh}
\end{figure}

The use of electron-phonon Wannier interpolation allows us to investigate
the sensitivity of the superconducting gap to the electron and phonon meshes
used for the calculations. Figure~\ref{gap0_mesh} shows the energy 
distribution of the $\sigma$-gap at $T$=30~K for eight sets of electron and 
phonon meshes. If we compare the gaps shown in Fig.~\ref{gap0_mesh} and the 
average coupling strengths reported in Table~\ref{table1} we see that 
obtaining converged results for $\Delta_{\bbk}$ is considerably more 
challenging than for  $\lambda$. For example, when using the same 
$\bbk$-points mesh ($80^3$) and different $\bbq$-points meshes ($20^3$ and 
$40^3$), the spread of the superconducting gap distribution changes from 
$\simeq$2.5~meV to $\simeq$1.5~meV, while the average coupling strength 
$\lambda$ is essentially unaffected. The same observation applies when we 
compare results for the same $\bbq$-points mesh ($40^3)$ but different 
$\bbk$-points meshes ($40^3$ and $80^3$). These differences highlight the 
difficulty of describing anisotropic quantities, and point out the need of 
having a very dense sampling of the Brillouin zone not only for the electrons 
but also for the phonons. For this reason the combination of the Eliashberg 
formalism with electron-phonon Wannier interpolation demonstrated in this work 
provides an ideal computational tool for investigating anisotropic 
superconductors.

\section{CONCLUSIONS}\label{section.conclusions}

In summary we developed a computational method which combines the 
anisotropic Migdal-Eliashberg formalism with electron-phonon interpolation
based on maximally-localized Wannier functions ({\tt EPW}). Our new method 
allows us to calculate the momentum- and band-resolved superconducting gap 
both effectively and accurately using a very fine description of 
electron-phonon scattering processes on the Fermi surface. In order to 
demonstrate our methodology we reported a comprehensive set of tests on two 
representative superconductors, namely Pb and MgB$_2$, and validated our 
approach against previous first-principles calculations as well as experiment. 
We discussed the performance of two analytic continuation methods for 
obtaining the superconducting gap on the real energy axis, and we
investigated the sensitivity of the calculated gaps on the underlying choice
of the Brillouin-zone grids for electrons and phonons.

In order to set a road map for first-principles studies of superconductors we 
now discuss the key approximations involved in the present approach and 
suggest possible avenues for future developments. The main approximations 
leading to Eqs.~(\ref{Znorm_surf}),(\ref{Delta_surf}) are as follows: (i) 
vertex corrections in the diagrammatic expansion of the electron 
self-energy are neglected following Migdal's theorem, (ii) the self-energy is 
assumed to be diagonal in band indices, (iii) the Eliashberg equations are 
restricted to bands near the Fermi level, (iv) the density of states near 
the Fermi level is assumed to be constant, and (v) the Coulomb interaction 
is described by an empirical 
parameter.~\cite{allen_mitrovic,pickett82,allen76} 

Regarding approximation (i), it is well known that Migdal's theorem can
break down for large electron-phonon coupling or in the presence of 
Fermi-surface nesting.~\cite{allen_mitrovic,pickett82} In this area the 
availability of electron-phonon matrix elements at a very small 
computational cost, as provided by our method, could be used as a starting 
point to explore the effects of vertex corrections beyond Migdal's 
approximation. For example the evaluation of the first non-crossing diagram 
should not constitute a major challenge, at least in the normal state. This 
may help assessing the numerical error introduced by Migdal's approximation.

Going beyond approximation (ii) is computationally challenging since
the Green's function and the self-energy should be treated as matrices in the
band indices.~\cite{garland,allen78} This step would increase substantially
the complexity of the formalism since the inversion of the Dyson's equation
would require matrix operations. Given the small energies associated with
the superconducting gap it is reasonable to expect that off-diagonal
terms should not play an important role in simple cases. However in the 
presence of degeneracies, and in particular in Jahn-Teller systems, the
correct description of these terms may prove critical.~\cite{gunnarsson}

It should be possible, at least in principle, to remove approximation (iii) by
including bands away from the Fermi level in the calculations. Here the 
difficulty is purely on the computational side, and for simple systems this 
should be doable. However relaxing this constraint would also introduce one 
additional equation [Eq.~(\ref{shift})] in order to calculate the correct 
quasiparticle shifts and impose the conservation of charge in the system.

The assumption (iv) of a constant density of states near the Fermi level
obviously breaks down for materials exhibiting structure in the density of 
states on the scale of the phonon energy. In theses cases it has been shown 
that the energy dependence of density of states can be retained within the 
isotropic approximation to the Eliashberg 
equations.~\cite{allen_mitrovic,pickett80,pickett82} We are planning to 
include this possibility in our methodology in the future.  

Lastly, approximation (v) means that in the present implementation the Coulomb 
repulsion between electrons remains described at the empirical level as an 
adjustable parameter. In order to remove this limitation our first step will
be to evaluate the Coulomb parameter using the dielectric matrix in the 
random-phase approximation.~\cite{cohen} In the longer term it would be 
desirable to introduce inside Eq.~(\ref{Delta_surf}) matrix elements of the 
screened Coulomb interaction calculated using the Sternheimer-GW method of 
Ref.~\onlinecite{sgw}.

We hope that the method reported here will prove useful to the 
superconductivity community as a robust and rigorous procedure for shedding 
light on existing superconductors and possibly for predicting new 
superconductors yet to be discovered.

\acknowledgments

E.R.M. was funded by Marie Curie IEF project FP7-PEOPLE-2009-IEF-252586. F.G. 
acknowledges support from the European Research Council under the EU FP7/ERC 
grant no. 239578.


\begin{thebibliography}{99}

\bibitem{carbotte}
J. P. Carbotte, Rev. Mod. Physics {\bf 62}, 1027 (1990). 

\bibitem{allen_mitrovic}
P. B. Allen, and B. Mitrovi\'{c}, Solid State Phys. {\bf 37}, 1 (1982).

\bibitem{marsiglio_book}
F. Marsiglio, and J. P. Carbotte, in {\it Superconductivity}, edited by 
K. H. Bennemann, and J. B. Ketterson (Springer 2008), Vol. 1, page 73.

\bibitem{scalapino66}
D. J. Scalapino, J. R. Schrieffer, and J. W. Wilkins, Phys. Rev. {\bf 148}, 
263 (1966).

\bibitem{scalapino}
D. J. Scalapino in {\it Superconductivity}, edited by R. D. Parks 
(Dekker, New York, 1969), Vol. 1, page 449.

\bibitem{scdft1} 
M. L\"{u}ders, M. A. L. Marques, N. N. Lathiotakis, A. Floris, G. Profeta, 
L. Fast, A. Continenza, S. Massidda, and E. K. U. Gross, 
Phys. Rev. B {\bf 72},  024545 (2005).

\bibitem{scdft2}
M. A. L. Marques, M. L\"{u}ders, N. N. Lathiotakis, G. Profeta, A. Floris, 
L. Fast, A. Continenza, E. K. U. Gross, and S. Massidda, 
Phys. Rev. B {\bf 72}, 024546 (2005).

\bibitem{bcs}
J. Bardeen, L. N. Cooper, and J. R. Schrieffer, Phys. Rev. {\bf 108}, 1175 
(1957). 
 
\bibitem{mcmillan}
W. L. McMillan, Phys. Rev. {\bf 167}, 331 (1968).

\bibitem{migdal}
A. B. Migdal, Zh. Eksperim. i Teor. Fiz. {\bf 34}, 1438 (1958) [Sov. Phys.
JETP {\bf 7}, 996 (1958)].

\bibitem{eliashberg}
G. M. Eliashberg, Zh. Eksperim. i Teor. Fiz. {\bf 38}, 966 (1960) [Sov.
Phys. JETP {\bf 11}, 696 (1960)].

\bibitem{golubov}
A. A. Golubov, J. Kortus, O. V. Dolgov, O. Jepsen, Y. Kong,
O. K. Andersen, B. J. Gibson, K. Ahn, and R. K. Kremer, J. Phys.: Condens. 
Matter {\bf 14}, 1353 (2002). 

\bibitem{choi_2band}
H. J. Choi, M. L. Cohen, and S. G. Louie, Phys. Rev. B {\bf 73}, 104520 (2006).

\bibitem{mazin}
I. I. Mazin, O. K. Andersen, O. Jepsen, A. A. Golubov, O. V. Dolgov, and 
J. Kortus, Phys. Rev. B {\bf 69}, 056501 (2004).

\bibitem{floris_mgb2}
A. Floris,  G. Profeta, N. N. Lathiotakis, M. L\"{u}ders, M. A. L. Marques, C. 
Franchini, E. K. U. Gross, A. Continenza, and S. Massidda, Phys. Rev. Lett. {\bf 94}, 037004 (2005). 

\bibitem{sanna12}
A. Sanna, S. Pittalis, J. K. Dewhurst, M. Monni, S. Sharma, G. Ummarino, 
S. Massidda, and E. K. U. Gross, Phys. Rev. B {\bf 85}, 184514 (2012). 

\bibitem{choi_mgb2}
H. J. Choi, D. Roundy, H. Sun, M. L. Cohen, and S. G. Louie, Nature (London) 
{\bf 418}, 758 (202); H. J. Choi, D. Roundy, H. Sun, M. L. Cohen, and S. G. 
Louie, Phys. Rev. B {\bf 66} , 020513(R) (2002).

\bibitem{sanna07}
A. Sanna, G. Profeta, A. Floris, A. Marini, E. K. U. Gross, and
S. Massidda, Phys. Rev. B {\bf 75}, 020511(R) (2007).

\bibitem{giustino_graphane}
G. Savini, A. C. Ferrari, and F. Giustino, Phys. Rev. Lett. {\bf 105}, 037002 
(2010).

\bibitem{kolmogorov_lib1}
A. N. Kolmogorov and S. Curtarolo, Phys. Rev. B {\bf 73}, 180501(R) (2006).

\bibitem{kolmogorov_lib2}
A. N. Kolmogorov, M. Calandra, and S. Curtarolo, Phys. Rev. B {\bf 78}, 094520 
(2008).

\bibitem{kolmogorov_feb}
A. N. Kolmogorov, S. Shah, E. R. Margine, A. F. Bialon, T. Hammerschmidt, and 
R. Drautz, Phys. Rev. Lett. {\bf 105}, 217003 (2010).

\bibitem{ceder}
A. Jain, G. Hautier, C. J. Moore, S. P. Ong, C. C. Fischer, T. Mueller, K. A. 
Persson, and G. Ceder, Computational Materials Science {\bf 50}, 2295 (2011).

\bibitem{curtarolo}
K. Yang, W. Setyawan, S. Wang, M. Buongiorno-Nardelli, and S. Curtarolo,
Nature Materials {\bf 11}, 614 (2012). 

\bibitem{arpes}
A. Damascelli, Z. Hussain, Z.-X. Shen, Rev. Mod. Phys. {\bf 75}, 473 (2003) 

\bibitem{sts}
O. Fischer, M. Kugler, I. Maggio-Aprile, C. Berthod, and C. Renner, Rev. Mod. 
Phys. {\bf 79}, 353 (2007).

\bibitem{giustino_wannier}
F. Giustino, M. L. Cohen, and S. G. Louie, Phys. Rev. B {\bf 76}, 165108 
(2007).

\bibitem{marzari2012}
N. Marzari, A. A. Mostofi, J. R. Yates, I. Souza, and D. Vanderbilt, Rev. Mod. 
Phys. {\bf 84}, 1419 (2012). 

\bibitem{yates}
J. R. Yates, X. Wang, D. Vanderbilt, and I. Souza, Phys. Rev. B {\bf 75}, 
195121 (2007). 

\bibitem{souza}
X. Wang, J. R. Yates, I. Souza, and D. Vanderbilt, Phys. Rev. B {\bf 74}, 
195118 (2006).

\bibitem{vanderbilt}
D. R. Hamann and D. Vanderbilt, Phys. Rev. B {\bf 79}, 045109 (2009).

\bibitem{EPW}
J. Noffsinger, F. Giustino, B. D. Malonea, C.-H. Park, S. G. Louie, and M. L. 
Cohen, Comput. Phys. Commun. {\bf 181}, 2140 (2010).

\bibitem{nambu}
Y. Nambu, Phys. Rev. {\bf 117}, 648 (1960).

\bibitem{gorkov}
L. P. Gorkov, Sov. Phys. JETP {\bf 7}, 505 (1958).

\bibitem{pickett82}
W. E. Pickett, Phys. Rev. B {\bf 26}, 1186 (1982).

\bibitem{allen76}
P. B. Allen, Phys. Rev. B {\bf 13}, 1416 (1976).

\bibitem{Anderson}
P. Morel, and P. W. Anderson, Phys. Rev. B {\bf 125}, 1263 (1962).

\bibitem{moon}
C.-Y. Moon, Y.-H. Kim, and K. J. Chang, Phys. Rev. B {\bf 70}, 104522 (2004).

\bibitem{liu_Pb}
A. Y. Liu and A. A. Quong, Phys. Rev B {\bf 53}, R7575 (1996).

\bibitem{bauer}
J. Bauer, J. E. Han, and O. Gunnarsson, arXiv:1202.5051v2. 

\bibitem{holcomb}
M. J. Holcomb, Phys. Rev. B {\bf 54}, 6648 (1996). 

\bibitem{vidberg}
H. J. Vidberg, and J. W. Serene, J. Low Temp. Phys. {\bf 29}, 179 (1977).

\bibitem{leavens}
C. R. Leavens, and D. S. Ritchie, Solid State Commun. {\bf 53}, 137 (1985). 

\bibitem{marsiglio}
F. Marsiglio, M. Schossmann, and J. P. Carbotte, Phys. Rev. B {\bf 37}, 4965 
(1988). 

\bibitem{beach}
K. S. D. Beach, R. J. Gooding, and F. Marsiglio, Phys. Rev. B {\bf 61}, 5147 
(2000).

\bibitem{baroni2001}
S. Baroni, S. de Gironcoli, A. Dal Corso, and P. Giannozzi, Rev. Mod. Phys. 
{\bf 73}, 515 (2001).

\bibitem{marzari1997}
N. Marzari and D. Vanderbilt, Phys. Rev. B {\bf 56}, 12847 (1997). 

\bibitem{souza2001}
I. Souza, N. Marzari, and D. Vanderbilt, Phys. Rev. B {\bf 65}, 035109 (2001). 

\bibitem{QE}
P. Giannozzi {\it et al.}, J. Phys. Condens. Matter {\bf 21}, 395502 (2009).

\bibitem{WANNIER}
A. A. Mostofi, J. R. Yates, Y.-S. Lee, I. Souza, D. Vanderbilt, and N. 
Marzari, Comput. Phys. Comm. {\bf 178}, 685 (2008).

\bibitem{allen-dynes}
P. B. Allen and R. C. Dynes, Phys. Rev. B {\bf 12}, 905 (1975).

\bibitem{broyden}
C. G. Broyden, Math. Comput. {\bf 19}, 577 (1965).

\bibitem{johnson}
D. D. Johnson, Phys. Rev. B {\bf 38}, 12807 (1988).

\bibitem{lda1}
D. M. Ceperley and B. J. Alder, Phys. Rev. Lett. {\bf 45}, 566 (1980).

\bibitem{lda2}
J. P. Perdew and A. Zunger, Phys. Rev. B {\bf 23}, 5048 (1981).

\bibitem{nc1}
N. Troullier and J. L. Martins, Phys. Rev. B {\bf 43}, 1993 (1991).

\bibitem{nc2}
M. Fuchs and M. Scheffler, Comput. Phys. Commun. {\bf 119}, 67 (1999).

\bibitem{mp}
M. Methfessel and A. T. Paxton, Phys. Rev. B {\bf 40}, 3616 (1989).

\bibitem{Pb_exp}
W. L. McMillan and J. M. Rowell, in {\it Superconductivity}, edited by
R. D. Parks (Dekker, New York, 1969), Vol 1, page 601.

\bibitem{blackford}
B. L. Blackford and R. H. March, Phys. Rev. {\bf 186}, 397 (1969).

\bibitem{lykken}
G. I. Lykken, A. L. Geiger, K. S. Dy, and E. N. Mitchell, Phys. Rev. B 
{\bf 4}, 1523 (1971).

\bibitem{tomlinson_Pb}
P. G. Tomlinson and J. P. Carbotte, Phys. Rev B {\bf 13}, 4738 (1976).

\bibitem{floris_Pb}
A. Floris, A. Sanna, S. Massidda, and E. K. U. Gross, Phys. Rev B {\bf 75}, 
054508 (2007).

\bibitem{dalcorso}
A. Dal Corso, J. Phys.: Condens. Matter {\bf 20}, 445202 (2008).

\bibitem{heid}
R. Heid, K.-P. Bohnen, I. Yu. Sklyadneva, and E. V. Chulkov, Phys. Rev B {\bf 
81}, 174527 (2010).

\bibitem{MgB2_exp}
J. Nagamatsu, N. Nakagawa, T. Muranaka, Y. Zenitani, and J. Akimitsu, Nature 
(London) {\bf 410}, 63 (2001).

\bibitem{kortus}
J. Kortus, I. I. Mazin, K. D. Belashchenko, V. P. Antropov, and L. L. Boyer,
Phys. Rev. Lett. {\bf 86}, 4656 (2001).

\bibitem{liu_mgb2}
A. Y. Liu, I. I. Mazin, and J. Kortus, Phys. Rev. Lett. {\bf 87}, 087005 
(2001).

\bibitem{giubileo}
F. Giubileo, D. Roditchev, W. Sacks, R. Lamy, D. X. Thanh, J. Klein, S. 
Miraglia, D. Fruchart, J. Marcus, and Ph. Monod, Phys. Rev. Lett. {\bf 87}, 
177008 (2001). 

\bibitem{bohnen}
K.-P. Bohnen, R. Heid, and B. Renker, Phys. Rev. Lett. {\bf 86}, 5771 (2001). 

\bibitem{eiguren}
A. Eiguren and C. Ambrosch-Draxl, Phys. Rev. B {\bf 78}, 045124 (2008).

\bibitem{calandra}
M. Calandra, G. Profeta, and F. Mauri, Phys. Rev. B {\bf 82}, 165111 (2010).

\bibitem{iavarone}
M. Iavarone, G. Karapetrov, A. E. Koshelev, W. K. Kwok, G. W. Crabtree, D. G. 
Hinks, W. N. Kang, E.-M. Choi, H. J. Kim, H.-J. Kim, and S. I. Lee, Phys. Rev. 
Lett. {\bf 89}, 187002 (2002). 

\bibitem{szabo}
P. Szab\'{o}, P. Samuely, J. Ka\v{c}mar\v{c}\'{i}k, T. Klein, J. Marcus, D. 
Fruchart, S. Miraglia, C. Marcenat, and A. G. M. Jansen, Phys. Rev. Lett. {\bf 
87}, 137005 (2001).

\bibitem{gonnelli}
R. S. Gonnelli, D. Daghero, G. A. Ummarino, V. A. Stepanov, J. Jun, S. M. 
Kazakov, and J. Karpinski, Phys. Rev. Lett. {\bf 89}, 247004 (2002).

\bibitem{garland}
J. C. Garland, Phys. Rev. {\bf 153}, 460 (1967).  

\bibitem{allen78}
P. B. Allen, Phys. Rev. B {\bf 18}, 5217 (1978). 

\bibitem{gunnarsson}
O. Gunnarsson, Rev. Mod. Phys. {\bf 69}, 575 (1997). 

\bibitem{pickett80}
W. E. Pickett, Phys. Rev. B {\bf 21}, 3897 (1980). 

\bibitem{cohen}
K.-H. Lee, K. J. Chang, and M. L. Cohen, Phys. Rev. B {\bf 52}, 1425 (1995). 

\bibitem{sgw}
F. Giustino, M. L. Cohen, and S. G. Louie, Phys. Rev. B {\bf 81}, 115105 (2010).

\end{thebibliography}
\end{document}